\documentclass[]{iopart}
\usepackage{graphicx}
\usepackage{iopams}
\usepackage{color}
\usepackage{bm} 
\usepackage{bbm, dsfont} 
\usepackage{subfigure}
\usepackage{mathrsfs}
\usepackage{verbatim}
\usepackage{cite}

\begin{document}

\title{Quantum-coherence-enhanced subradiance in a chirally coupled atomic chain}

\author{H. H. Jen}
\address{Institute of Physics, Academia Sinica, Taipei 11529, Taiwan}
\address{Institute of Atomic and Molecular Sciences, Academia Sinica, Taipei 10617, Taiwan}
\ead{sappyjen@gmail.com} 

\renewcommand{\k}{\mathbf{k}}
\renewcommand{\r}{\mathbf{r}}
\newcommand{\parallelsum}{\mathbin{\!/\mkern-5mu/\!}}
\def\p{\mathbf{p}} 
\def\R{\mathbf{R}}
\def\bea{\begin{eqnarray}}
\def\eea{\end{eqnarray}}
\date{\today}
\begin{abstract} 
We theoretically study the quantum-coherence-enhanced subradiance in a chiral-coupled atomic chain with nonreciprocal decay channels. The collective radiation in this one-dimensional (1D) nanophotonics system results from the resonant dipole-dipole interactions in 1D reservoirs, which allow infinite-range couplings between atoms. When single photon interacts with part of the atomic chain from a side excitation, the subradiant decay can be further reduced when highly correlated states are initially excited. The excitation plateau in the decay process can emerge due to the ordered population exchanges, which presents one distinctive signature of long-range and light-induced atom-atom correlations. Multiple time scales of the decay behaviors also show up due to multiple scattering of light transmissions and reflections in the chain. We further investigate the effect of atomic position fluctuations, and find that the cascaded scheme with a uni-directional coupling is more resilient to the fluctuations, while the overall decay constant can be increased due to large deviations. Our results present a fundamental study on the subradiance and light-induced atom-atom correlations in such 1D nanophotonics platforms, and offer rich opportunities in potential applications of quantum storage of photons.
\end{abstract}

\flushbottom

\section{Introduction}

One-dimensional (1D) nanophotonics systems \cite{Chang2018} recently raise many interests in the capability to manipulate strong and infinite-range light-matter couplings \cite{Solano2017}. This infinite-range coupling originates from resonant dipole-dipole interactions (RDDI) in 1D reservoirs \cite{Tudela2013}, an extension to RDDI in free space \cite{Lehmberg1970}, which emerge due to photon rescattering in the radiative process. In such 1D atom-fiber or atom-waveguide systems, strong coupling regime can be fulfilled to initiate superradiance \cite{Dicke1954, Gross1982} from the guided photons that enable multiple scattering of light within the atoms \cite{Goban2015}. Moreover, chiral quantum optics \cite{Lodahl2017} can be realized in such 1D nanophotonics setups, which allows nonreciprocal decay channels that break the time-reversal symmetry. The mechanism behind this chiral-coupled interface is due to the evanescent waves \cite{Bliokh2014, Bliokh2015} at the glass-air surface under total internal reflection, where spin(light polarization)-momentum(light propagation direction) locking \cite{Lodahl2017} emerges, such that particular polarized light can only propagate to its correlated direction.  

In such 1D nanophotonics systems with effective chiral couplings, an atom-fiber system shows directional spontaneous emissions controlled by initialized atomic internal states \cite{Mitsch2014}, and quantum dot in waveguides can form an interface of spin qubit and path-encoded photons \cite{Luxmoore2013} or implement a Mach-Zehnder interferometer \cite{Sollner2015}. Many-body dimerized states \cite{Stannigel2012, Ramos2014, Pichler2015} are theoretically proposed by engineering the nonreciprocal decay channels of the chiral-coupled atomic chain, where emerging universal dynamics can be demonstrated under the coherent part of the system \cite{Kumlin2018}. Recently, chiral quantum link in distant two atomic arrays in free-space is proposed to enable quantum state transfer \cite{Grankin2018}, and strong photon-photon correlations in a 1D chiral-coupled atomic ensemble are also predicted \cite{Mahmoodian2018}. With the flexibility of tuning nonreciprocal couplings and the scalability of solid-state platforms, 1D nanophotonics systems can provide an alternative quantum interface of controlling light-matter interactions, and allow potential applications in many-body state preparations \cite{Pichler2015}, spin dynamics simulation \cite{Hung2016}, and selective transport of atomic excitations \cite{Jen_transport_2019}. A recent progress on generating single collective excitation \cite{Corzo2019} in an atom-nanofiber system enables an alternative interface to process quantum information and paves way toward quantum network \cite{Kimble2008} in a strong coupling regime. 

\begin{figure}[t]
\centering
\includegraphics[width=10cm,height=7.5cm]{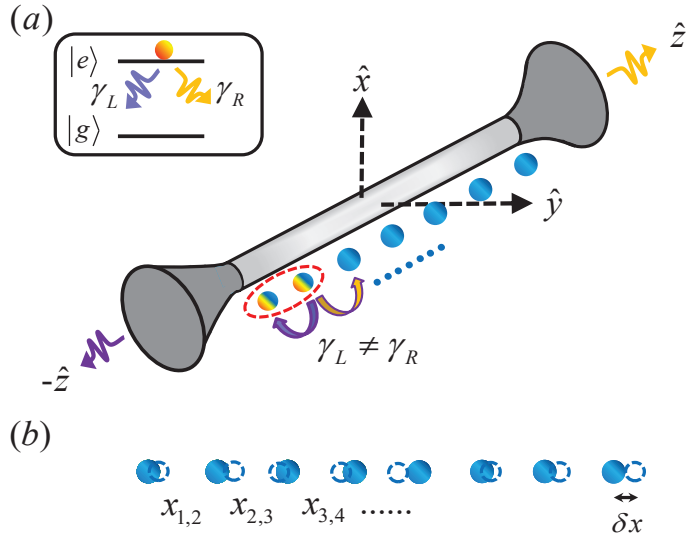}
\caption{Schematic chiral-coupled atomic chain with single excitation. (a) A one-dimensional atom-fiber coupled system demonstrates an effective chiral coupling with nonreciprocal decay channels of $\gamma_L\neq\gamma_R$ along $\hat z$. Two-level quantum emitters ($|g(e)\rangle$ indicating the ground and excited states respectively) are aligned along $\hat z$, where two atoms denoted by a dashed ellipse, as an example, are uniformly excited by single photon from side excitation along $\hat x$. The effective chiral-coupled atomic chain guides the decaying photon throughout the whole chain via these nonreciprocal decay channels. (b) The atomic chain with inter-atomic spacings of $x_{1,2}$, $x_{2,3}$, $x_{3,4}$, and so on, can be manipulated to control the light-matter interactions, while the atoms can be subjected to position fluctuations $\delta x$ (dashed circles).}\label{fig1}
\end{figure}

In this article, we investigate the subradiance in the chiral-coupled atomic chain as shown in Figure \ref{fig1}. In addition to the subradiant modes measured in plasmonic ring nanocavities \cite{Sonnefraud2010} and ultracold molecules \cite{McGuyer2015}, the subradiant state preparations \cite{Scully2015, Facchinetti2016, Jen2016_SR, Jen2016_SR2, Sutherland2016, Bettles2016, Jen2017_MP, Garcia2017, Plankensteiner2017, Guimond2019} and light scattering \cite{Guerin2016, Bromley2016, Zhu2016, Shahmoon2017, Jenkins2017, Jen2018_SR1, Jen2018_SR2} are mostly studied in free-space dense atoms, while the subradiance is less explored in such 1D chiral-coupled systems. Here we consider a single-photon excitation on part of the atomic chain, and show that the subradiance initiated by the subradiant coupling of 1D RDDI can be enhanced by quantum coherences of the initially excited states. The rest of the paper is organized as follows. We first introduce the coupled equations for a 1D chiral-coupled atomic chain with single excitation. Next we characterize the subradiance property for various nonreciprocal decay channels and number of atoms. We further study the effect of position fluctuations of the atoms on the dissipation, and finally we give a conclusion.

\section{Effective theoretical model of chiral couplings}

In free space, the radiative decay of the excited atom is isotropically emitted. This shows the nature of reciprocal system-reservoir interaction which initiates the spontaneous emission. When many atoms are involved, significant RDDI emerge due to photon rescattering within the dense medium. The RDDI couple every other atoms in pairs and are symmetric when any two atoms exchange their positions. This again originates from the preservation of the time reversal symmetry in the system-reservoir interaction. By contrast, in the atom-fiber or atom-waveguide system we consider in Figure \ref{fig1}, this 1D reservoir allows infinite-range couplings in sinusoidal forms \cite{Tudela2013, Kumlin2018}, which can be structured \cite{Scelle2013, Chen2014, Ramos2014} for nonreciprocal decay channels. The effective chiral master equations for such 1D atom-light interacting system reads \cite{Pichler2015}, 
\begin{eqnarray}
\frac{d \rho}{dt}=-\frac{i}{\hbar}[H_L+H_R,\rho]+\mathcal{L}_L[ \rho]+\mathcal{L}_R[ \rho],\label{rho}
\end{eqnarray}
which involves the left (L)- and right (R)-coupling terms respectively. The coherent parts are
\begin{eqnarray}
H_L\equiv-\frac{i\hbar\gamma_L}{2}\sum_{\mu<\nu}\left(e^{ik|x_\mu-x_\nu|}\sigma_\mu^\dag\sigma_\nu-\textrm{H.c.}\right),\\
H_R\equiv-\frac{i\hbar\gamma_R}{2}\sum_{\mu>\nu}\left(e^{ik|x_\mu-x_\nu|}\sigma_\mu^\dag\sigma_\nu-\textrm{H.c.}\right),
\end{eqnarray}
and the Lindblad forms are defined as
\begin{eqnarray}
\mathcal{L}_L[\hat \rho]\equiv-\frac{\gamma_L}{2}\sum_{\mu,\nu}^N\Big\{e^{-ik(x_\mu-x_\nu)}\left(\sigma_\mu^\dag \sigma_\nu \rho +\rho \sigma_\mu^\dag\sigma_\nu -2\sigma_\nu \rho\sigma_\mu^\dag\right)\Big\},\\
\mathcal{L}_R[\hat \rho]\equiv-\frac{\gamma_R}{2}\sum_{\mu,\nu}^N\Big\{e^{ik(x_\mu-x_\nu)}\left(\sigma_\mu^\dag \sigma_\nu \rho +\rho \sigma_\mu^\dag\sigma_\nu -2\sigma_\nu \rho\sigma_\mu^\dag\right)\Big\},
\end{eqnarray}
where $k=2\pi/\lambda$ is the wave vector for the transition wavelength $\lambda$, $\sigma_\mu^\dag\equiv|e\rangle_\mu\langle g|$, and $\sigma_\mu=(\hat\sigma_\mu^\dag)^\dag$ with $|g(e)\rangle$ for the ground and excited state respectively. $\gamma_{L(R)}$ characterizes the left- and right-propagating decay rates. In the above Lindblad forms, we do not include the non-guided or non-radiative losses, which can mitigate the efficiency of light collections via fibers or waveguides.

The usual reciprocal and infinite-range couplings can be retrieved when $\gamma_L=\gamma_R=\gamma$, which take the sinusoidal forms of
\begin{eqnarray}
J_{\mu,\nu}=\gamma\left[\cos(k_L x_{\mu,\nu})+i\sin(k_L |x_{\mu,\nu}|)\right],
\end{eqnarray}
after combining the above $H_L+H_R$ and $\mathcal{L}_L[\hat \rho]+\mathcal{L}_R[\hat \rho]$ respectively. Re(Im)[$J_{\mu,\nu}$] represents the dissipative(coherent) parts respectively, and $x_{\mu,\nu}\equiv x_\mu-x_\nu$. And as such, equation (\ref{rho}) can be reduced to 
\begin{eqnarray}
\frac{d \rho}{dt}=-\frac{i}{\hbar}[\rm{Im}(J_{\mu,\nu}),\rho]-\rm{Re}(J_{\mu,\nu})\left(\sigma_\mu^\dag \sigma_\nu \rho +\rho \sigma_\mu^\dag\sigma_\nu -2\sigma_\nu \rho\sigma_\mu^\dag\right).
\end{eqnarray}
This collective and infinite-range dipole-dipole interaction in the 1D atom-fiber coupled system has been investigated theoretically \cite{Kien2005, Kien2008, Kien2017} and is recently observed between two clouds near a nanofiber separated by several hundreds of transition wavelengths \cite{Solano2017}. 

When single photon interacts with the atomic chain, the Hilbert space of the system is limited to the ground $|g\rangle^{\otimes N}$ and singly-excited states $|\psi_\mu\rangle=\sigma_\mu^\dag|g\rangle^{\otimes N}$. This is the weak excitation limit considered in the coherent dipole model \cite{Bromley2016, Zhu2016, Sutherland2016_2} or low saturation regime used in Green's function approach \cite{Garcia2017_2}. Within a single-excitation space, the system dynamics after single photon absorption \cite{Scully2006} can be described by 
\begin{eqnarray}
|\Psi(t)\rangle=\sqrt{1-\sum_{\mu=1}^N|A_\mu(t)|^2}|g\rangle^{\otimes N}+\sum_{\mu=1}^N A_\mu(t)|\psi_\mu\rangle,
\end{eqnarray}
where the probability amplitude $A_\mu(t)$ can be obtained by the coupled equations,
\begin{eqnarray}
\dot{A}_\mu(t)=\sum_{\nu=1}^N V_{\mu,\nu}A_\nu(t),\label{A}
\end{eqnarray}
with the given initial conditions of $\vec A(t=0)$ where $\vec A\equiv$ $[A_1(t),$ $A_2(t),$ $...,A_N(t)]$. The chiral-coupled interaction $V$ composed of matrix elements $V_{\mu,\nu}$ can be obtained from equation (\ref{rho}) under a single-excitation space, 
\begin{eqnarray}
V_{\mu,\nu}=\left\{\begin{array}{lr}
    -\gamma_Le^{-ik|x_{\mu,\nu}|},~\mu<\nu\\
		-\frac{\gamma_L+\gamma_R}{2}\delta_{\mu,\nu}\\
		-\gamma_Re^{-ik|x_{\mu,\nu}|},~\mu>\nu
\end{array}\right.,\label{V}
\end{eqnarray}
where the nonsymmetric feature emerges in exchanging the atomic indices $\mu$ and $\nu$ when $\gamma_R \neq \gamma_L$, while $V$ becomes reciprocal, that is $V_{\mu,\nu}=V_{\nu,\mu}$, only when $\gamma_L=\gamma_R$. In general $V$ is not a normal matrix since $VV^\dag\neq V^\dag V$, and furthermore it is a defective matrix which can not be eigen-decomposed in terms of linearly independent eigenvectors. Therefore, a method of singular value decomposition using the left and right eigenvectors is not able to study the non-Hermitian $V$ with nonsymmetric matrix elements. Below we directly solve the coupled equations in time evolutions, and investigate the subradiant dynamics of the chiral-coupled atomic chain.

\section{Quantum-coherence-enhanced subradiance}

In the chiral-coupled atomic chain with the effective coupling in equation (\ref{V}), three main system parameters determine the radiative properties of the single photon excitation. The first is the directionality [$D=(\gamma_R-\gamma_L)/(\gamma_R+\gamma_L)$] \cite{Mitsch2014} which defines how much radiation propagates toward the right over the left. In essence, this also characterizes the amount of light transmissions and reflections within the atomic chain, where light transfer is carried out via atomic deexcitation and reabsorption. The other is the inter-atomic spacings $k |x_{\mu,\nu}|$, which specify the pairwise and infinite-range couplings in the atomic chain. In this section, we consider an equidistant atomic array, and as such the effect of the pairwise coupling relies only on $e^{-im\xi}$ with $m\equiv |\mu-\nu|$ for any two atoms and $\xi\equiv k |x_{\mu,\mu+1}|$. To investigate the subradiant dynamics in particular, we consider the parameter regime of $\xi=\pi$, which attributes to the decoherence-free subspace when the system is uniformly excited under reciprocal couplings ($\gamma_L=\gamma_R=\gamma$). This can be seen in the eigenvalues of equation (\ref{V}) for $N=2$ with reciprocal couplings, which leads to two eigen-decay constants of $\gamma(e^{-i\xi}-1,-e^{-i\xi}-1)$ for an arbitrary $\xi$. When $\xi=2n\pi$ with integers $n$, these constants correspond respectively to the eigenvectors $(\mp|eg\rangle+|ge\rangle)/\sqrt{2}$ in a single-excitation space, which are exactly the singlet (anti-symmetric) and triplet (symmetric) states in Dicke's bases \cite{Dicke1954}. In this strong coupling regime, the triplet state gives the maximal decay rate of $2\gamma$, showing superradiance with an enhanced rate proportional to the number of atoms. On the other hand, the decoherence-free state emerges in the triplet subspace when $\xi=\pi$.

Furthermore, the configurations of initial atomic excitations present another crucial element in describing the dissipation process after single photon absorption. The atomic excitations determine the initial quantum coherence of the system, which depends on how many atoms are interacting with this photon. This will determine how the light-induced atom-atom correlations, for example of $\langle\sigma_{\mu\neq\nu}^\dag\sigma_\nu\rangle$, build up as the photon propagates throughout the medium, which manifests in the subradiance dynamics after excitation. Below we consider a side single-photon excitation either on the end or the central part of the atomic chain, and investigate the role of initial quantum coherence in the radiative properties.
 
\subsection{End excitations}

Here we study the effect of initial atomic excitations on the subradiance in the chiral-coupled atomic chain with $\xi=\pi$. Since the system is one-dimensional, we can order the atoms as $x_1<x_2<...<x_N$. When single photon interacts with $N_i$ atoms starting from the end of the chain, from a side excitation as proposed in Figure \ref{fig1}, the system forms a W state on absorption, which we denote it as the initialized state of the system,
\begin{eqnarray}
|\Psi(t=0)\rangle=\frac{1}{\sqrt{N_i}}\sum_{\mu=1}^{N_i}\sigma_\mu^\dag|g\rangle^{\otimes N}.
\end{eqnarray} 
The W state is known for the maximally entangled state with a dimension of $N_i$. We vary $N_i$ to control the initial quantum coherence and study its effect on the dissipation or transport of the atomic excitations through the atomic chain. When $N_i$ increases, more atoms are correlated initially, and below we will show that this modifies the subradiance property significantly. To prepare such highly entangled states \cite{Corzo2019}, the single photon can be focused or guided by single mode fibers to excite $N_i$ atoms specifically. 

We note that in $|\Psi(t=0)\rangle$, there are no correlated phases between the excited atoms due to the side excitation at a right angle to the axis of the atomic chain. For other kinds of initialized states, for example of the states with correlated phases from the excitation, we may expect a regime of enhanced superradiant emissions. Here, we mainly focus on the subradiance property which can be enhanced via controlling the coherences of W states, and therefore we leave a different regime of superradiance for further study in the future. 

\subsubsection{Cascaded scheme}

First we consider the cascaded scheme \cite{Stannigel2012, Gardiner1993, Carmichael1993} where $\gamma_R=\gamma$ and $\gamma_L=0$. This presents the case with a uni-directional decay channel, which does not allow backward light transfer, and thus the reflection of light is forbidden. Since the left decay channel vanishes, we are able to obtain the iterative expression for various probability amplitudes $A_\mu(t)$ in equation (\ref{A}). For $N_i$ atoms with arbitrary $\xi$ with the initial conditions of $A_{\mu\leq N_i}(0)=1/\sqrt{N_i}$, we have
\begin{eqnarray}
A_{m+1}(t)=-e^{-\frac{\gamma t}{2}-im\xi}\int_0^t dt\sum_{m'=1}^m A_{m'}(t) e^{\frac{\gamma t}{2}+i(m'-1)\xi},
\end{eqnarray}
where $m\geq N_i$. The above form presents only one exponential function of $e^{-\gamma t/2}$, which originates from the intrinsic decay of $\gamma_R/2$ for individual atoms. 

Take $N_i=1$ as an example, we show some results of $A_m(t)$ for small $m\leq 4$,
\begin{eqnarray}
&&A_1(t)=e^{-\gamma t/2},\\
&&A_2(t)=-te^{-\gamma t/2-i\xi},\\
&&A_3(t)=\frac{1}{2}t(t-2)e^{-\gamma t/2-i2\xi},\\
&&A_4(t)=-\frac{1}{6}t(t^2-6t+6)e^{-\gamma t/2-i3\xi}.
\end{eqnarray}
Under this particular initialized state of $N_i=1$, which is unentangled and no light-induced atom-atom correlation is present in the beginning, various excited state populations $P_m(t)\equiv |A_m(t)|^2$ do not depend on $\xi$. The population of the first atom $P_1(t)$ decays purely exponentially as in the independent case without dipole-dipole interactions. This happens in the cascaded scheme where the leftmost atom has no feedback coupling from the atoms on the right, whereas the rest of the atoms can be repopulated via the deexcitation of the atoms on the left. Interestingly, $A_{m>2}(t)$ of the $m$th atom involves $(m-2)$ zero points other than $t=0$, which reflects the sign change in its probability amplitude, leading to the repopulation. The distribution of these zero points of $t^{(m)}_1$, $t^{(m)}_2$, $\cdots$, $t^{(m)}_{m-2}$ exactly follows the order of
\begin{eqnarray}
t^{(m+1)}_1<t^{(m)}_1<t^{(m+1)}_2<t^{(m)}_2<\cdots
<t^{(m+1)}_{m-2}<t^{(m)}_{m-2}<t^{(m+1)}_{m-1},\label{rule}
\end{eqnarray}        
for $A_m(t)$ and $A_{m+1}(t)$, respectively, which suggests the ordered oscillations and population exchanges between the nearest-neighbor atoms.

When $N_i\geq 2$, $\xi$ starts to play a role in $A_m(t)$. We take $N_i=2$ as an example again for arbitrary $\xi$, and we obtain $A_m(t)$ for small $m\leq 3$ as
\begin{eqnarray}
&&A_1(t)=\frac{e^{-\gamma t/2}}{\sqrt{2}},\\
&&A_2(t)=\frac{(e^{i\xi}-t)e^{-\gamma t/2-i\xi}}{\sqrt{2}},\\
&&A_3(t)=-\frac{t(2+2e^{i\xi}-t)e^{-\gamma t/2-i2\xi}}{2\sqrt{2}}.
\end{eqnarray}
When $\xi=2n\pi$, $A_2(t)\propto (1-t)$ and $A_3(t)\propto t(4-t)$, which decrease at the early stage of the decay, whereas $A_2(t)\propto (1+t)$ and $A_3(t)\propto t^2$ extend in time when $\xi=\pi$. This explains again why we particularly focus on the parameter of $\xi=\pi$, which puts the system into a subradiant regime. In this regime, we again have similar rule of equation (\ref{rule}) for the distribution of $(m-N_i-1)$ zero points of $A_{m>N_i+1}(t)$ in general.         

\begin{figure}[t]
\centering
\includegraphics[width=10cm,height=5.2cm]{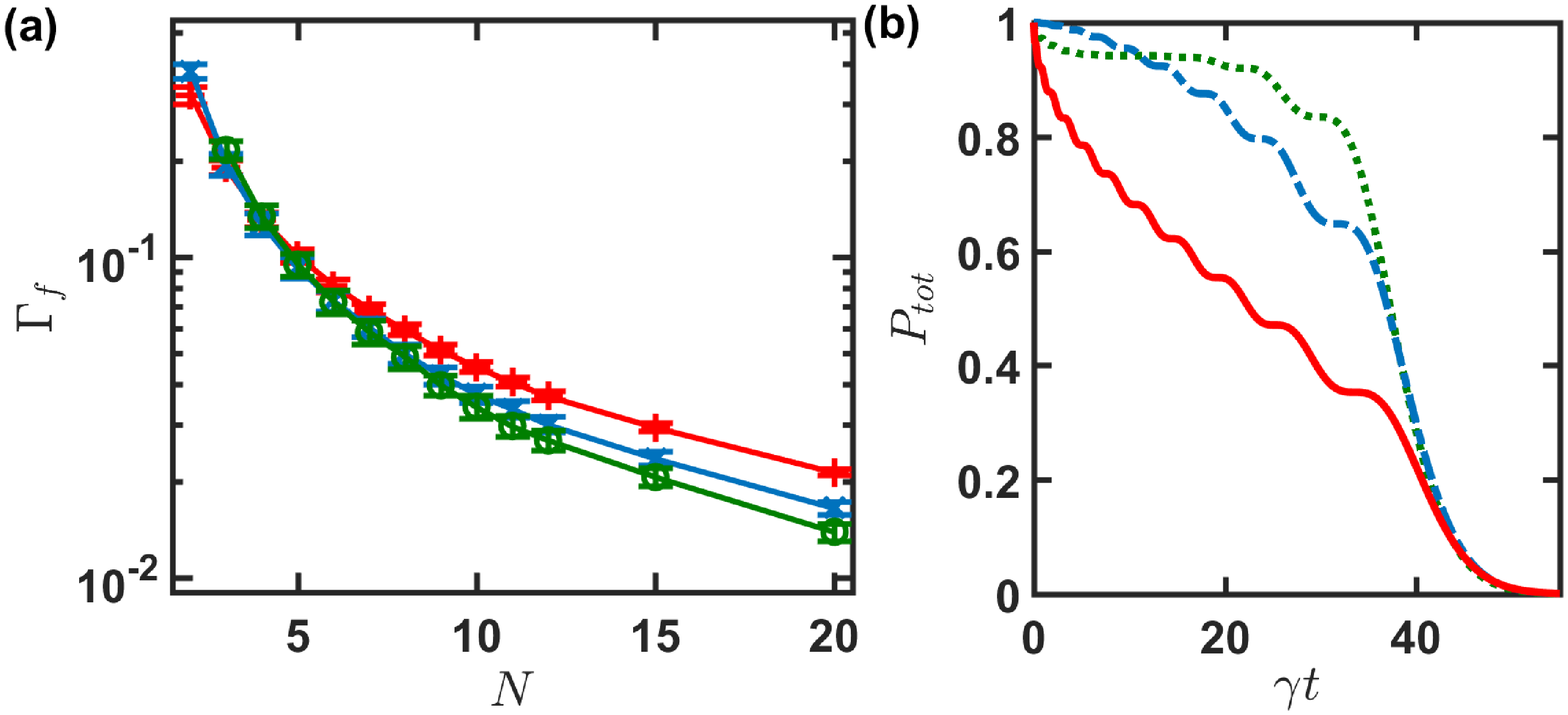}
\caption{Effective decay constant and total excited state populations in the cascaded scheme for $\gamma_R=\gamma$ and $\gamma_L=0$, at $\xi=\pi$. (a) The effective decay constants $\Gamma_f$ are obtained by fitting $P_{tot}$ with an exponentially decay function $e^{-\Gamma_f t}$ down to $10^{-3}$ of $P_{tot}(t=0)$. We consider various atomic excitations of $N_i=1$ ($+$), $2$ ($\times$), and $3$ ($\circ$) starting from the end of the atomic chain with a total number of atoms $N\geq N_i$. Larger $N_i$ represents more initialized correlated atoms on absorption of single photon, which leads to a more subradiant decay. The error bars denote a $95\%$ confidence level of the fitted $\Gamma_f$. (b) Time evolutions of the total excited state populations for $N=12$ with $N_i=1$ (solid-red), $2$ (dashed-blue), and $3$ (dotted-green).}\label{fig2}
\end{figure}

To characterize the general decay behaviors for different $N_i$, in Figure \ref{fig2}(a) we plot the effective decay constants $\Gamma_f$, which we obtain by fitting the total excited state populations $P_{tot}\equiv\sum_{\mu=1}^N |A_\mu(t)|^2$ with an exponential function $e^{-\Gamma_f t}$. This gives an overall time scale for given initialized atomic excitations. As more atoms share the single photon absorption and form highly correlated W states in the beginning, the effective decay rate decreases, which indicates that the system supports a more subradiant emission. Moreover, some overlaps of $\Gamma_f$ for small $N$ indicate the boundary effect of the atomic chain. And as such, for $N=2$, the case of $N_i=2$ can decay faster than $N_i=1$, similarly for $N=3$ with $N_i=3$ and $N_i=2$ respectively. This boundary effect weakens as $N\gtrsim 2N_i$, where light-induced correlations have enough time to build up throughout the whole atomic chain before the photon leaves completely. Larger $N_i$ predominantly presents a lower $\Gamma_f$, which demonstrates an enhanced subradiance due to significant quantum coherence within the initial atomic excitations.

In Figure \ref{fig2}(b), we show the total excited state population which becomes even out as $N_i$ increases, along with small regions of excitation plateaus. It is this flattened region that prolongs the overall decay time of the population. We leave the explanations to the next section of the detailed subradiance dynamics related to the excitation plateaus. Meanwhile, the overall decay does not behave quite as an exponentially decaying function, especially for larger $N_i$. We note that at least two decay behaviors emerge before and after $\gamma t\sim 40$ in Figure \ref{fig2}(b). For $\gamma t\lesssim 40$, more significant subradiance appears for a larger $N_i$, where a power-law decay in this range of time better describes the dissipation process. Nevertheless, we still use the fitted $\Gamma_f$ as an estimate to characterize the overall decay, and as a comparison to other cases of non-cascade scheme and different excitation configurations below. We will show that multiple time scales of the excitation decays emerge, and therefore $\Gamma_f$ can serve as a good macroscopic measure in various parameter regimes.  

\subsubsection{Non-cascaded scheme}

Next we turn on a finite decay channel of $\gamma_L$ and study the subradiance dynamics in the chiral-coupled atomic chain under the non-cascaded scheme. A finite decay channel of $\gamma_L$ allows more significant quantum interference within the atomic chain. This manifests most significantly when $\gamma_L=\gamma_R$ in the strong coupling regime of $\xi=n\pi$, where eigenvalues of $-\gamma(N,0,0,\cdots,0)$ present $(N-1)$ highly degenerate decoherence-free modes and one superradiant mode with a decay constant of $N\gamma$. This extreme case of strong coupling regime to initiate Dicke's superradiance turns out to be possible in 1D chiral-coupled atomic chain due to its infinite-range nature of RDDI. On the contrary in conventional free-space atomic systems, this strong coupling regime can not be reached due to strong dipole-dipole interaction energy shifts that increase significantly as $\xi$ becomes smaller \cite{Lehmberg1970}. 

\begin{figure}[t]
\centering
\includegraphics[width=10cm,height=5.2cm]{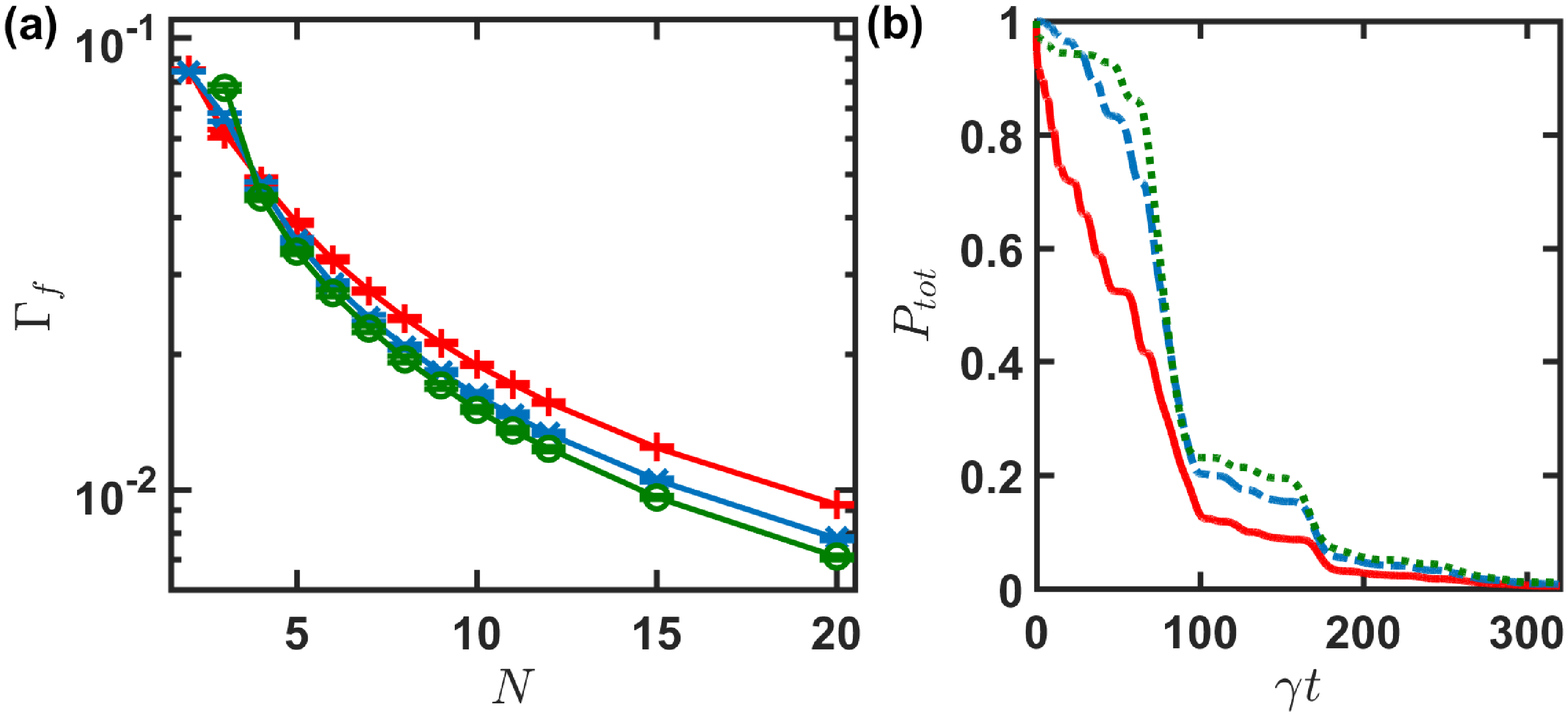}
\caption{Effective decay constant and total excited state populations in the non-cascaded scheme for $\gamma_R=\gamma$ and $\gamma_L=0.5\gamma_R$, at $\xi=\pi$. (a) The effective decay constants $\Gamma_f$ are obtained as in Figure \ref{fig2} for $N_i=1$ ($+$), $2$ ($\times$), and $3$ ($\circ$). Larger $N_i$ again leads to a more subradiant decay, but each case of $N_i$ has smaller $\Gamma_f$ than the cascaded scheme of Figure \ref{fig2} for the same number of atoms $N$. (b) Time evolutions of the total excited state populations for $N=12$ with $N_i=1$ (solid-red), $2$ (dashed-blue), and $3$ (dotted-green).}\label{fig3}
\end{figure}

As a comparison to Figure \ref{fig2}, we demonstrate the non-cascaded scheme with a moderate $\gamma_L$ in Figure \ref{fig3}. Similar to the boundary effects observed in Figure \ref{fig2}(a), the fitted $\Gamma_f$ in Figure \ref{fig3}(a) is larger for $N\lesssim 2N_i$, whereas it shows an enhanced subradiance when $N\gtrsim 2N_i$ for an increasing $N_i$. We note that the scale of $\Gamma_f$ in the non-cascaded scheme is much smaller than the cascaded case since light can exchange between the atoms via either transmission or reflection, and such that the photon behaves as if trapped in the chain. In Figure \ref{fig3}(b), the total excited state populations present another distinctive feature on the multiple scalings of time constants in the decay. Before and after $\gamma t\sim 100$ and $200$ respectively, three distinguishing drops of the populations can be identified, in contrast to Figure \ref{fig2}(b) with only two obvious separations of time scales. These differences also reflect on the excitation plateaus which span over a broader range of time in the non-cascaded scheme between $\gamma t=100$ and $200$ in Figure \ref{fig3}(b). By contrast the plateaus in Figure \ref{fig2}(b) have smaller ranges, similar to the early stage of Figure \ref{fig3}(b) before $\gamma t\sim 100$. This suggests that the broader excitation plateau can only be enabled when significant subradiant decay is permitted under a finite nonreciprocal decay channel $\gamma_L$.  

\begin{figure}[t]
\centering
\includegraphics[width=10cm,height=5.2cm]{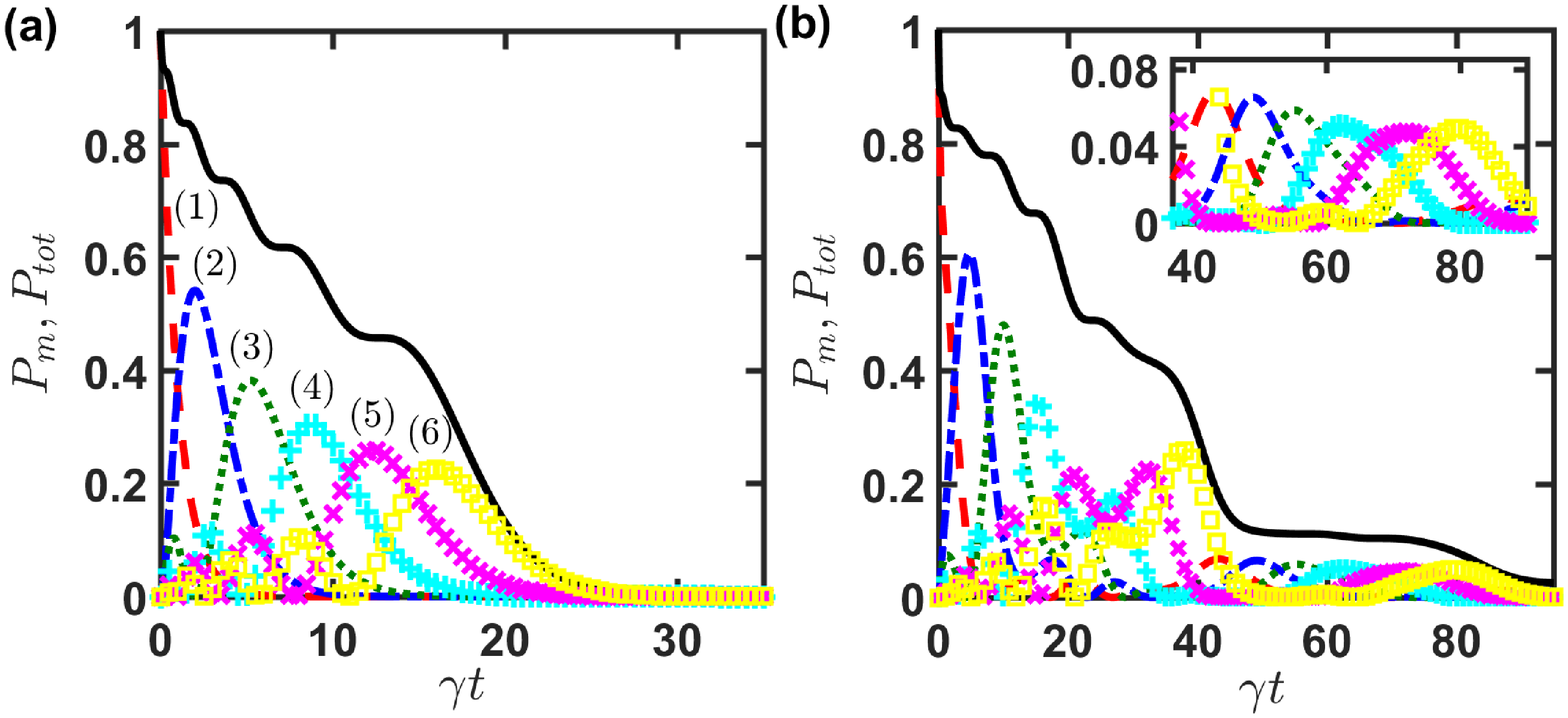}
\caption{Subradiant dynamics at $\xi=\pi$ for $N=6$ with $N_i=1$. Time evolutions of $P_{tot}$ (solid-black) and excited state populations $P_m$ for individual atoms denoted by the parentheses are plotted in (a) the cascaded scheme of $\gamma_R=\gamma$ and $\gamma_L=0$ and (b) the non-cascaded scheme of $\gamma_R=\gamma$ and $\gamma_L=0.5\gamma_R$, respectively. Excitation plateaus can be seen as signature of ordered atomic excitations $P_m$. The inset of (b) is a zoom-in of an even broader excitation plateau between $\gamma t\sim 40-80$.}\label{fig4}
\end{figure}

To unravel different decay behaviors for cascaded and non-cascaded scheme, we show the detailed subradiance dynamics for individual atoms in Figure \ref{fig4}. We choose a smaller $N=6$ and $N_i=1$ as an example. The cascaded scheme in Figure \ref{fig4}(a) presents an ordered atomic excitations, where each excitation plateau can be formed approximately between neighboring atoms during successive population exchange. On the other hand, the non-cascaded scheme in Figure \ref{fig4}(b) shows an even broader plateau at $\gamma t\gtrsim 40$. Smaller excitation plateaus also appear at $\gamma t\lesssim 20$, similar to Figure \ref{fig4}(a). A larger excitation plateau takes time to show up, which indicates a finite time of establishment of correlations, and therefore it can only be evident in a more subradiant parameter regime. We note that the zeros of numerically calculated $P_{m}(t)$ in Figure \ref{fig4}(a) indeed distribute according to the rule obtained in equation (\ref{rule}), which corresponds to ordered population exchanges between neighboring atoms. 

\begin{figure}[h]
\centering
\includegraphics[width=10cm,height=5.2cm]{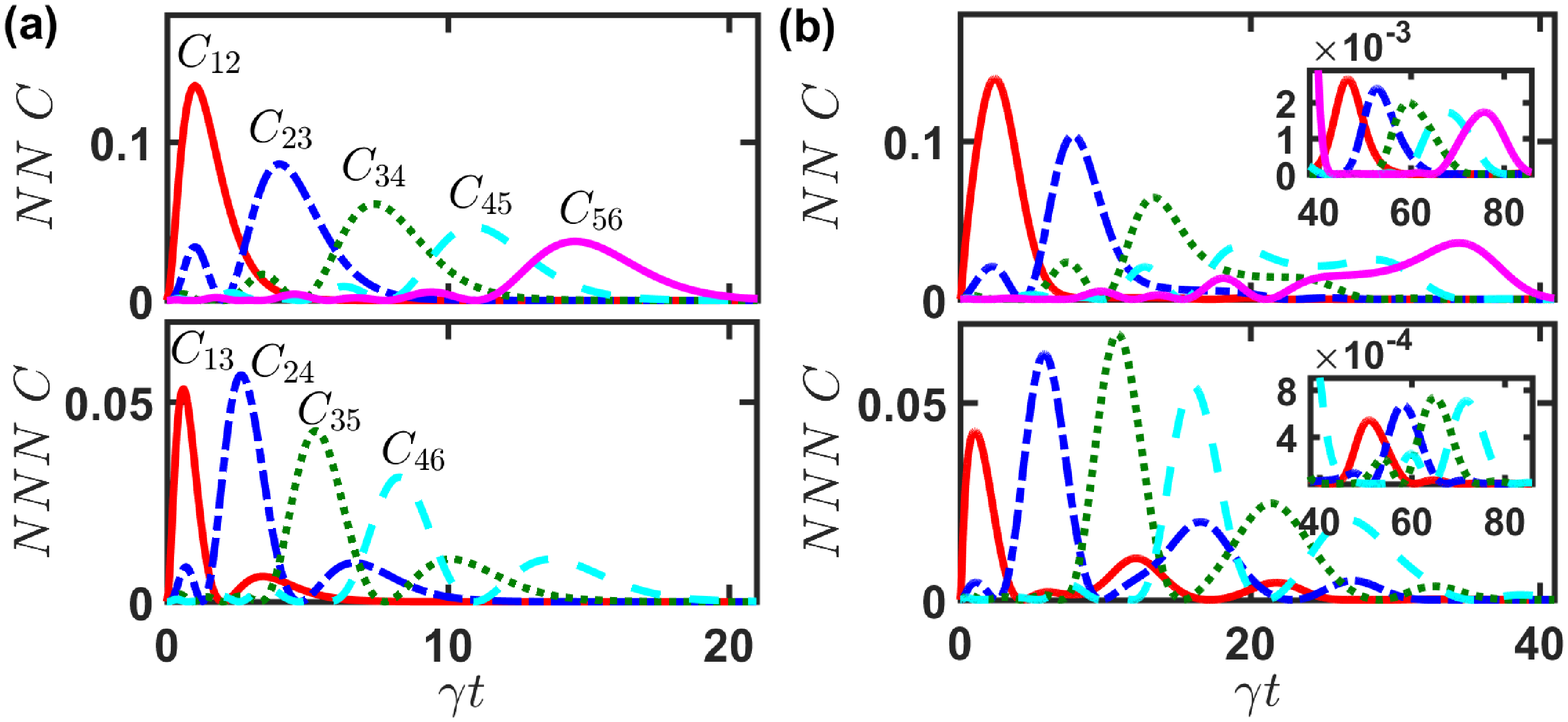}
\caption{Light-induced atom-atom correlations for $N=6$ and $N_i=1$. We plot nearest-neighbor (NN) and next NN correlations (C) for the (a) cascaded and (b) non-cascaded scheme, corresponding to the same parameters of Figure \ref{fig4}. We denote various atom-atom correlations by $C_{\mu\nu}\equiv|\langle\sigma_\mu^\dag\sigma_\nu\rangle|^2$ in the plots. Finite NN and NNN correlations are shown in the insets of (b), indicating correlated atomic excitations at longer time and corresponding to the broader excitation plateau in Figure \ref{fig4}(b).}\label{fig4_2}
\end{figure}

These small excitation plateaus can be regarded to significant nearest-neighbor (NN) excitation correlations. Furthermore, the broader plateau suggests of more correlated atomic excitations, which presents an emerging long-range correlation. This can be seen in the insets of Figure \ref{fig4}(b), where finite NN and next NN (NNN) correlated excitations are present. We identify these correlations in Figure \ref{fig4_2} in more details, and obtain $C_{\mu\nu}\equiv|\langle\sigma_\mu^\dag\sigma_\nu\rangle|^2$ under the same the parameters of Figure \ref{fig4}. In contrast to relatively well separated NN correlations in the upper plot of Figure \ref{fig4_2}(a), the non-cascaded scheme shows prolonged NN correlations of $C_{45}$ and $C_{56}$ as shown in the upper plot of Figure \ref{fig4_2}(b), which marks the onset of the establishment toward atom-atom correlations at longer distances. We further plot the next NN correlations, and find that $C_{13 (24)}$ and $C_{35 (46)}$ in the non-cascaded scheme are within the same correlation envelopes respectively (near $\gamma t\sim 10(15)$), in contrast to the ones in the lower plot of Figure \ref{fig4_2}(a). This is a manifestation of even longer-range correlations of $C_{15 (26)}$ in the non-cascaded scheme. In the insets of Figure \ref{fig4_2}(b), we look more closely at a longer time, where these correlations overlap more with each other and thus enable a broader excitation plateau in Figure \ref{fig4}(b). 

\subsection{Central excitations}

Here we further study a different excitation configuration where central part of the atomic chain is singly excited. The cascaded scheme for central excitations has no difference from the end excitations since the atoms only couple each other with one-way decay channel. Therefore for end excitations with an odd $N$ and $N_i$, the central excitations present the same subradiance dynamics for an atomic chain with the same $N_i$ but different total $(2N-N_i)$ atoms. 

\begin{figure}[t]
\centering
\includegraphics[width=10cm,height=5.2cm]{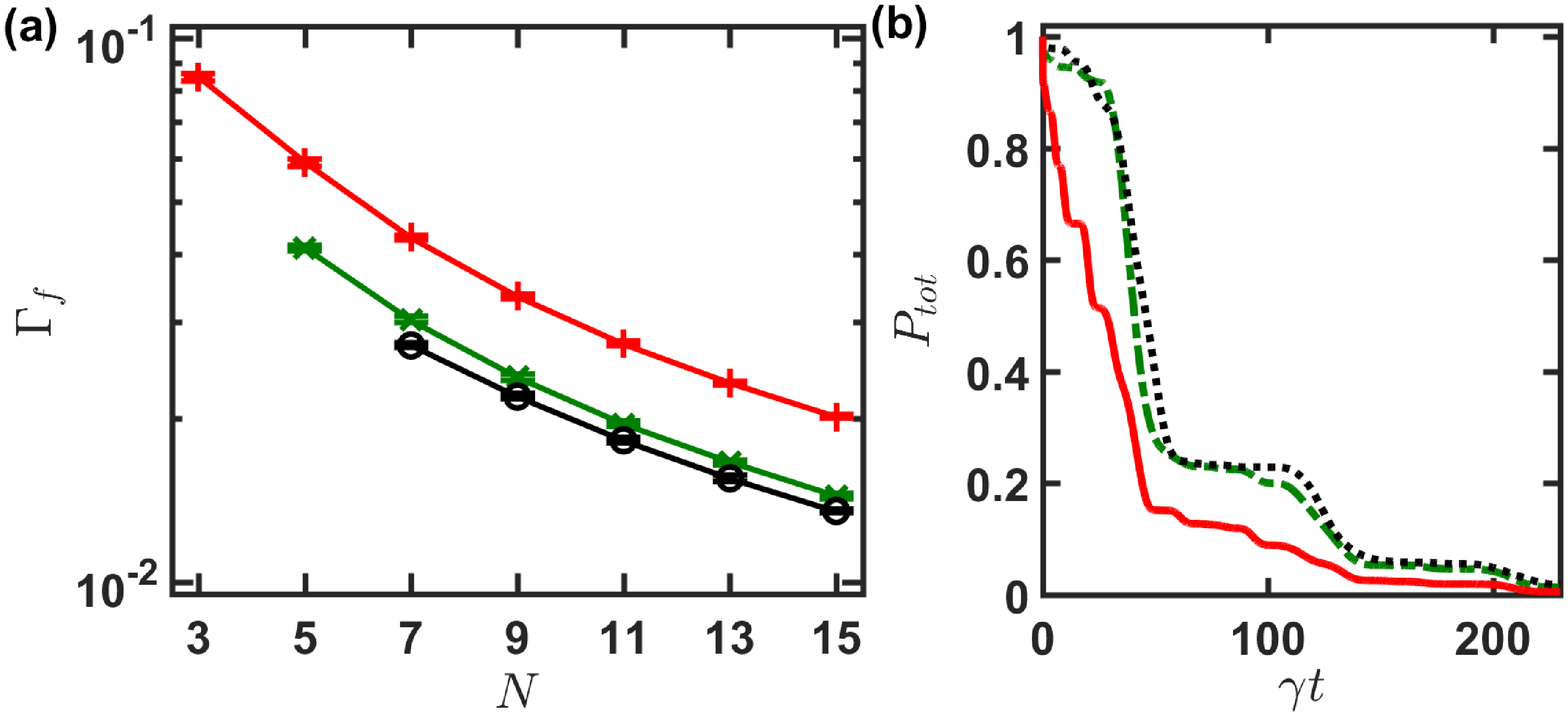}
\caption{Effective decay constant and total excited state populations in the non-cascaded scheme with central excitation. As a comparison, we choose the same $\gamma_R=\gamma$ and $\gamma_L=0.5\gamma_R$ as in Figure \ref{fig3}. (a) The effective decay constants $\Gamma_f$ for $N_i=1$ ($+$), $3$ ($\times$), and $5$ ($\circ$). Larger $N_i$ leads to a more subradiant decay, but not as significant as in Figure \ref{fig3} for the same number of atoms $N$. (b) Time evolutions of the total excited state populations for $N=11$ with $N_i=1$ (solid-red), $3$ (dash-dotted-green), and $5$ (dotted-black).}\label{fig5}
\end{figure}

As a comparison, we consider the same non-cascaded scheme as in Figure \ref{fig3}. To reduce the boundary effect on the subradiant decay, we choose $N\geq N_i+2$ with an odd $N$. As shown in Figure \ref{fig5}(a), the $\Gamma_f$ shows quantum-coherence-enhanced subradiance when $N_i$ increases, similar to the end excitation configuration in Sec. 3.1. Furthermore, the fitted decay rate for $N_i=5$ almost saturates with the one of $N_i=3$ for the same $N$, which indicates that the subradiant decay can not be made smaller unlimitedly by using even more correlated W states initially. The $\Gamma_f$ in Figure \ref{fig5}(a) comparing the ones of the same $N$ in Figure \ref{fig3}(a) is larger, indicating a less enhanced subradiance. This can be explained by the effective number of participating atoms along the direction of the dominant decay channel, which determines how many photon exchanges and atomic deexcitation or repopulation engage in the dissipation. The configuration of central excitations has less effective participating atoms, compared to the end excitations which in principle involve the whole chain. Therefore, for the same $N$, the initial correlation of atomic excitation has less effect in enhancing the subradiance under the central excitations. To see whether this argument is reasonable, we take the $\Gamma_f$ for $N=15$ with $N_i=1$ and $3$ in Figure \ref{fig5}(a) as an example, which has effective number of participating atoms of $8$ and $9$ respectively. These $\Gamma_f$'s are approximately equal to the cases of $N=9$ and $10$ with $N_i=1$ and $3$ respectively in Figure \ref{fig3}(a), making the effective atom number a good measure of comparing subradiance behaviors. We note that as $N_i\rightarrow N$ or $N\rightarrow\infty$, the decay behaviors of both excitation configurations should approach each other. 

The saturation of $\Gamma_f$ as $N_i$ increases can be also seen in Figure \ref{fig5}(b), where $P_{tot}$ almost overlaps for $N_i=3$ and $5$. Similar to Figure \ref{fig3}(b) and Figure \ref{fig4}(b), $P_{tot}$ here shows broader excitation plateaus around $\gamma t\sim 100$, indicating again multiple scales of decay time and emerging long-range and light-induced atom-atom correlations.  

\section{Effect of atomic position fluctuations}

\begin{figure}[t]
\centering
\includegraphics[width=10.0cm,height=5.6cm]{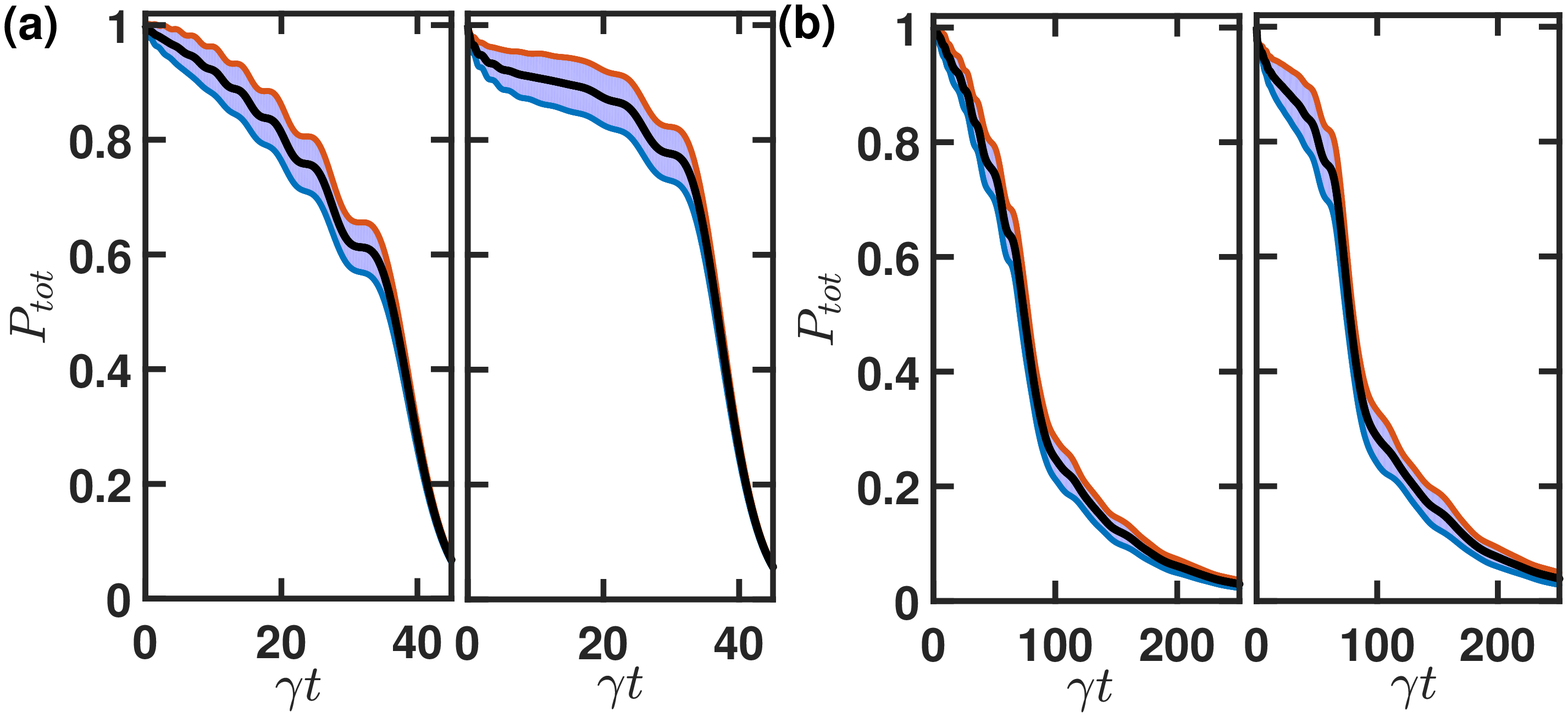}
\caption{Total excited state populations under position fluctuations for $N=12$. The position fluctuations are introduced to the (a) cascaded ($\gamma_R=\gamma$ and $\gamma_L=0$) and (b) non-cascaded ($\gamma_R=\gamma$ and $\gamma_L=0.5\gamma_R$) schemes, respectively with $20\%$ and $2\%$ randomly distributed deviations around the fixed position of $\xi=\pi$. The left and right panels in each plots (a) and (b) denote the initial excitations of $N_i=2$ and $3$ respectively. Shaded areas are filled between the upper and lower curves with $1\sigma$ standard deviation, and a solid-black line presents the mean value after converging ensemble averages.}\label{fig6}
\end{figure}

Finally we study the effect of position fluctuations on the subradiance dynamics in the chiral-coupled atomic chain. We include this effect to better compare with realistic experiments, which should have a pronounced effect on neutral atoms with optical transitions, but less, for example, on the superconducting qubits with microwave transmissions. 

In Figure \ref{fig6}, we introduce a fraction of position fluctuations relative to $\xi$ on each atoms of the chain. We choose initial atomic excitations of $N_i=2$ and $3$ respectively for the cascaded and non-cascaded schemes, and compare the results with position fluctuations to Figures \ref{fig2} and \ref{fig3}. We note that the excited state population with $N_i=1$ in the cascaded scheme does not depend on $\xi$, as we have addressed in the section of cascaded scheme. In Figure \ref{fig6}(a) for the cascaded scheme, the excitation plateaus withstand the fluctuations up to $20\%$, and they start to smooth out when the fluctuations are more than $40\%$. On the contrary, in Figure \ref{fig6}(b), $2\%$ position fluctuation already has smeared out the plateau structure near $\gamma t\lesssim 100$ in the non-cascaded scheme. As the fluctuations increase, the cascaded scheme is more resilient to them compared to the non-cascaded scheme. This can be due to the fact that non-cascaded scheme allows both decay channels, and such that more notable effect of position fluctuations emerge via two-way couplings, whereas only uni-directional coupling is permitted in the cascaded scheme, which makes it less affected. Furthermore, $P_{tot}$ deviates more significantly than its mean value for a larger $N_i$, indicating that more correlated W states are more fragile under the position fluctuations. This also leads to the reduction of lifetime comparing to $P_{tot}$ without fluctuations, which is similar to the suppression of linewidth narrowing in the subradiant eigenmodes for a two-dimensional array \cite{Facchinetti2016} or a ring structure \cite{Jen2018_SR2} of quantum emitters with position fluctuations.


\section{Conclusion}

In conclusion, we have investigated the subradiant property of single photon excitation in the chirally coupled atomic chain. In the subradiant coupling regime, we consider a side excitation on part of the chain with equidistant separations, such that we can manipulate the initial quantum coherence on absorption of single photon. This initially excited and highly correlated W state presents a quantum-coherence-enhanced subradiance due to ordered atomic excitations along the chain, leading to multiple excitation plateaus in the decay. The excitation plateau further corresponds to long-range and light-induced atom-atom correlations due to the ordered population exchange. Moreover we show the emerging multiple time scales of the decay when multiple scattering of light transmissions and reflections are allowed under non-cascaded couplings in the chain. We finally introduce the effect of atomic position fluctuations on these subradiance properties. The cascaded scheme with uni-directional coupling is more resilient to the fluctuations, while the overall decay time can be reduced due to large deviations. We present a fundamental study on the subradiance dynamics in a chirally coupled chain, and demonstrate strong light-induced atom-atom correlations in such 1D nanophotonics platforms. This strongly interacting system can offer many opportunities in a potential application of photon storage \cite{Sayrin2015}, and provide a platform to simulate long-range quantum magnetism \cite{Hung2016}, to access the interaction-driven phases of bi-edge (hole) excitations \cite{Jen2020_PRR}, and to study disorder-induced localized excitations \cite{Jen2020_PRA} with tunable chiral and infinite-range couplings.

\ack

This work is supported by the Ministry of Science and Technology (MOST), Taiwan, under the Grant No. MOST-106-2112-M-001-005-MY3. We thank Y.-C. Chen, G.-D. Lin, and M.-S. Chang for insightful discussions, and are also grateful for NCTS ECP1 (Experimental Collaboration Program).

\end{document}